\documentclass{article}
\usepackage[latin9]{inputenc}
\usepackage{esint}
\usepackage{graphicx}
\usepackage{subfigure}
\usepackage{bm}
\usepackage{amsfonts}

\begin{document}

\title{BSW process of the slowly evaporating charged black hole}

\author{Liancheng Wang, Feng He and Xiangyun Fu}
\maketitle
\begin{abstract}
In this paper, we study the BSW process of the slowly evaporating
charged black hole. It can be found that the BSW process will also
arise near black hole horizon when the evaporation of charged black
hole is very slow. But now the background black hole does not have
to be an extremal black hole, and it will be approximately an extremal
black hole unless it is nearly a huge stationary black hole.
\end{abstract}

\section{Introduction}

Ba\"nados, Silk and West (BSW) process was first pointed out in \cite{banados_kerr_2009}
that a rotating black hole may accelerate a particle to arbitrarily
high energy when two particles collide near the horizon. As a remarkable
way to extract energy from blackholes, it evokes great interest and
argument\cite{banados_kerr_2009,abdujabbarov_particle_2013,banados_emergent_2011,berti_comment_2009,galajinsky_particle_2013,grib_particle_2011,harada_collision_2011-1,harada_collision_2011-2,harada_upper_2012,igata_effect_2012,jacobson_spinning_2010,lake_particle_2010,mcwilliams_black_2013,nakao_ultrahigh_2013,nemoto_escape_2013,patil_high_2011,patil_naked_2011,piran_high_1975,Tsukamoto2014,zaslavskii_comment_2013,zaslavskii_acceleration_2011,patil_acceleration_2012,wei_charged_2010,zaslavskii_acceleration_2009,Zhu:2011ja,Ding:2013nc,Liu:2010ja,Liu:2011wv,Yao:2011ai,kimura_acceleration_2011,Harada2011}.

Soon after the article \cite{banados_kerr_2009} appeared, BSW process
was extended to the charged black hole\cite{zaslavskii_acceleration_2009}
that particle accelerator may appear near the horizon of a charged
black hole without rotating. Not as the rotating black holes, only
few charged black holes have been studied\cite{patil_acceleration_2012,wei_charged_2010,zaslavskii_acceleration_2009,Zhu:2011ja}.
It may be an important reason that a charged black hole will reduce
to a stationary uncharged black hole rapidly.

Because of the existence of time variable, until now the BSW process
of the evaporating charged black hole has not been studied in literatures.
In this paper we will make a try to discuss this problem. For the
difficulty of time variable, we will avoid studying the fast evaporating
black hole, but turn to discuss the situation that the evaporation
of the charged black hole is very slow. \cite{kimura_acceleration_2011}
suggested that the extremal Kerr black hole is linearly unstable in
BSW process. In other words, if one of the test particles with an
arbitrarily small mass which can be described by a linear perturbation
has the fine-tuned angular momentum(or charge) such that the BSW process
works well, the center of mass(CM) energy will deviate perturbed state
and be magnified till as large as the mass energy of the background
rotating black hole (or the charged black hole). The occurrence of
this linearly unstability must satisfy two conditions: (1) The black
hole is extremal. (2) The particles have the correct angular momentums
(or charges). According to the above ideas, we describe the decay
of the charge as a small perturbation if the evaporation of the background
black hole is very slow. So we can deal with the collision process
of two particles by the usual method. Under this ansatz the BSW process
will occur just when the perturbation is invalid, i.e., the black
hole is linearly unstable, if the energy of the test particles matches
a correct value in the collision process.

This paper is organized as follows. The general geodesic motion and
the CM energy near the horizon of the slowly evaporating charged black
hole are given respectively in the following two sections. In section
4 we discuss the classification of critical particles in BSW process.
The last section is the conclusion of our paper.

\section{general geodesic motion}

The line element of the charged evaporating black hole is
\begin{equation}
ds^{2}=-e^{2\psi}f(r)dv^{2}+2e^{\psi}dvdr+r^{2}(d\theta^{2}+\sin^{2}\theta d\phi^{2}),\label{schw-1}
\end{equation}
where
\begin{equation}
f(r)=1-\frac{2M}{r}+\frac{Q^{2}}{r^{2}}.\label{fr-1}
\end{equation}
Here black holes mass $M=M(r)$, charge $Q=Q(r)$ and $\psi=\psi(r)$
are the hidefunction of $v$. The event horizon of the black hole
is

\begin{equation}
2e^{-\psi}\dot{r}=f(r),\label{eq:horizonEvap}
\end{equation}
where $\dot{}=\partial_{v},\,'=\partial_{r}$.

The Hamiltonian for the geodesic motion is
\begin{equation}
\mathcal{H}[x^{\alpha},p_{\beta}]=\frac{1}{2}g^{\mu\nu}p_{\mu}p_{\nu},\label{hamilton-1}
\end{equation}
where $p_{\mu}$ is the conjugate momentum to $x^{\mu}$. Let $S=S(\lambda,x^{\alpha})$
be the action of the parameter $\lambda$ and coordinates $x^{\alpha}$.
The conjugate momentum $p_{\alpha}$ is described by $p_{\alpha}=\partial S/\partial x^{\alpha}-qA$.
The Hamilton-Jacobi equation reads
\begin{equation}
-\frac{\partial S}{\partial\lambda}=\frac{1}{2}g^{\mu\nu}\left(\frac{\partial S}{\partial x^{\mu}}-qA_{\mu}\right)\left(\frac{\partial S}{\partial x^{\nu}}-qA_{\nu}\right),\label{hjequation-1}
\end{equation}
where $dA_{\nu}=-Q/rdv.$

For charged evaporating black hole, the action can be written as
\begin{equation}
S=\frac{1}{2}m^{2}\lambda-Ev+L\phi+S_{r}(r)+S_{\theta}(\theta),\label{sexpress-1}
\end{equation}
where $m$, $E$ and $L$ are the rest mass, energy and angular momentum
respectively. If the evaporation of black hole is very slow, $f\rightarrow0$(i.e.
$\dot{r}\rightarrow0$), we can separate variables by usual method.
Substituting Eq.(\ref{sexpress-1}) into Eq.(\ref{hjequation-1})
\begin{equation}
-m^{2}=2e^{-\psi}\left(-E+q\frac{Q}{r}\right)\frac{\partial S}{\partial r}+f\left(\frac{\partial S}{\partial r}\right)^{2}+\frac{1}{r^{2}}\left(\frac{\partial S}{\partial\theta}\right)^{2}+\frac{1}{r^{2}\sin^{2}\theta}L^{2},\label{hjmetric21-1}
\end{equation}
then separating variables and defining a constant $\mathcal{K}$
\begin{equation}
-m^{2}r^{2}-\left[2e^{-\psi}r^{2}\left(-E+q\frac{Q}{r}\right)\frac{\partial S}{\partial r}+fr^{2}\left(\frac{\partial S}{\partial r}\right)^{2}\right]=\mathcal{K}=\left(\frac{\partial S}{\partial\theta}\right)^{2}+\frac{1}{\sin^{2}\theta}L^{2},\label{hjmetric22}
\end{equation}
the Hamilton-Jacobi function becomes
\begin{equation}
S=\frac{1}{2}m^{2}\lambda-Ev+L\phi+\delta_{r}\int^{r}Rdr+\delta_{\theta}\int^{\theta}\sqrt{\Theta}d\theta,\label{hjr}
\end{equation}
where
\begin{eqnarray}
R & = & \frac{\partial S}{\partial r}=\frac{-e^{-\psi}\left(-E+q\frac{Q}{r}\right)\pm\sqrt{e^{-2\psi}\left(-E+q\frac{Q}{r}\right)^{2}-f\left(m^{2}+\frac{\mathcal{K}}{r^{2}}\right)}}{f},\label{hjrnote-1}\\
\Theta & = & \left(\frac{dS_{\theta}}{d\theta}\right)^{2}=\mathcal{K}-\frac{1}{\sin^{2}\theta}L{}^{2}.\label{hjthetanote-1}
\end{eqnarray}

With the aid of $p^{\alpha}=g^{\alpha\beta}p_{\beta}$, from Hamilton-Jacobi
Eq.(\ref{hjequation-1}), we can obtain the motion equations

\begin{eqnarray}
\frac{dv}{d\lambda} & = & e^{-\psi}\delta_{r}R,\\
\frac{dr}{d\lambda} & = & e^{-\psi}\left(-E+q\frac{Q}{r}\right)+f\delta_{r}R,\\
\frac{d\theta}{d\lambda} & = & \frac{1}{r^{2}}\sigma_{\theta}\sqrt{\Theta},\\
\frac{d\phi}{d\lambda} & = & \frac{1}{r^{2}\sin^{2}\theta}L,
\end{eqnarray}
and the effective potential is
\[
\sqrt{V}=e^{-\psi}\left(-E+q\frac{Q}{r}\right)+fR.
\]

\section{CM energy near the horizon}

The four momenta of two particles with rest masses $m_{1}$ and $m_{2}$
is $p_{(i)}^{\alpha}=m_{(i)}u_{(i)}^{\alpha}$, and the CM energy
$E_{CM}$ of the two particles defines
\begin{equation}
E_{CM}^{2}=m_{1}^{2}+m_{2}^{2}-2g^{ab}p_{(1)a}p_{(2)b}.\label{cmenergy-1}
\end{equation}
Hence the CM energy of two colliding general geodesic particles in
the evaporating charged black hole spacetime is
\begin{eqnarray}
E_{CM}^{2} & = & m_{1}^{2}+m_{2}^{2}-2[e^{-\psi}(-E_{1}+q_{1}\frac{Q}{r})\delta_{2r}R_{2}+e^{-\psi}(-E_{2}+q_{2}\frac{Q}{r})\delta_{1r}R_{1}\nonumber \\
 &  & +f\delta_{1r}R_{1}\delta_{2r}R_{2}+\frac{1}{r^{2}}\delta_{1\theta}\sqrt{\Theta_{1}}\delta_{2\theta}\sqrt{\Theta_{2}}+\frac{1}{r^{2}\sin^{2}\theta}L_{1}L_{2}].\label{cmenergytwo-1}
\end{eqnarray}

Unlike the stationary black hole, the evaporating black hole needs
an additional condition that the evaporation of the black hole is
very slow (i.e. $\dot{r}\rightarrow0$). Thus $f(r)\rightarrow0$
can been seen as a small parameter in the near-horizon limit. The
CM energy of two general geodesic particles in the near-horizon limit
can be written as
\begin{eqnarray}
E_{CM}^{2} & = & m_{1}^{2}+m_{2}^{2}\nonumber \\
 &  & -2\{-\frac{E_{2}r_{H}-q_{2}Q}{2(E_{1}r_{H}-q_{1}Q)}(m_{1}^{2}+\frac{\mathcal{K}_{1}}{r_{H}^{2}})-\frac{E_{1}r_{H}-q_{1}Q}{2(E_{2}r_{H}-q_{2}Q)}(m_{2}^{2}+\frac{\mathcal{K}_{2}}{r_{H}^{2}})\nonumber \\
 &  & +\frac{f(r)}{e^{-2\psi}}\frac{r_{H}^{2}}{8}[-\frac{(E_{2}r_{H}-q_{2}Q)}{(E_{1}r_{H}-q_{1}Q)^{3}}(m_{1}^{2}+\frac{\mathcal{K}_{1}}{r_{H}^{2}})^{2}\nonumber \\
 &  & -\frac{(E_{1}r_{H}-q_{1}Q)}{(E_{2}r_{H}-q_{2}Q)^{3}}(m_{2}^{2}+\frac{\mathcal{K}_{2}}{r_{H}^{2}})^{2}+\frac{2(m_{1}^{2}+\frac{\mathcal{K}_{1}}{r_{H}^{2}})(m_{2}^{2}+\frac{\mathcal{K}_{2}}{r_{H}^{2}})}{(E_{1}r_{H}-q_{1}Q)(E_{2}r_{H}-q_{2}Q)}]\nonumber \\
 &  & +\frac{1}{r_{H}^{2}}\delta_{1\theta}\sqrt{\Theta_{1}}\delta_{2\theta}\sqrt{\Theta_{2}}+\frac{1}{r_{H}^{2}\sin^{2}\theta}L_{1}L_{2}\}.\label{cmhorizonlimit-1}
\end{eqnarray}
We can see that the $E_{CM}$ in evaporating black hole is the same
as the $E_{CM}$ in stationary black hole\cite{zaslavskii_acceleration_2009}
if we omit the second order approximate term $f_{evaporate}r_{H}^{2}/8e^{-2\psi}$
in evaporating black hole (It is $f_{stationary}r_{H}^{2}/8$ in stationary
black hole). That is to say the evaporation of the charge has a little
effect on the CM energy, but we can see it will make a subtle effect
on the classification of critical particles from next section.

\section{classification of critical particles}

We denote

\begin{equation}
F=e^{-2\psi}(-E+q\frac{Q}{r}){}^{2}-f(m^{2}+\frac{\mathcal{K}}{r^{2}}),\label{eq:rright}
\end{equation}
then we have

\begin{equation}
F'=-f'\left(\frac{\mathcal{K}}{r^{2}}+m^{2}\right)+\frac{2\mathcal{K}f}{r^{3}}-2e^{-2\psi}\psi'\left(\frac{qQ}{r}-E\right)^{2}-\frac{2qQe^{-2\psi}\left(\frac{qQ}{r}-E\right)}{r^{2}}\label{eq:dprer}
\end{equation}
and

\begin{eqnarray}
F'' & = & -f''\left(\frac{\mathcal{K}}{r^{2}}+m^{2}\right)+\frac{4\mathcal{K}f'}{r^{3}}-\frac{6\mathcal{K}f}{r^{4}}\nonumber \\
 &  & +\frac{8qQe^{-2\psi}\psi'\left(\frac{qQ}{r}-E\right)}{r^{2}}+e^{-2\psi}\left(4\psi'^{2}-2\psi''\right)\left(\frac{qQ}{r}-E\right)^{2}\nonumber \\
 &  & +e^{-2\psi}\left[\frac{2q^{2}Q^{2}}{r^{4}}+\frac{4qQ\left(\frac{qQ}{r}-E\right)}{r^{3}}\right],\label{eq:ddprer}
\end{eqnarray}
where
\begin{eqnarray}
f' & = & -\frac{2M'}{r}+\frac{2M}{r^{2}}-\frac{2Q^{2}}{r^{3}}+\frac{2QQ'}{r^{2}},\label{dfdescribe-1}\\
f'' & = & -\frac{2M''}{r}+\frac{4M'}{r^{2}}-\frac{4M}{r^{3}}+\frac{6Q^{2}}{r^{4}}-\frac{8QQ'}{r^{3}}+\frac{2Q'^{2}+2QQ''}{r^{2}}.\label{ddfdescribe-1}
\end{eqnarray}

Below we will discuss the classification of critical particles by
using the idea of \cite{Harada2011}:

(1) $F=0$, it is

\[
e^{-2\psi}\left(-E+q\frac{Q}{r}\right)^{2}-f\left(m^{2}+\frac{\mathcal{K}}{r^{2}}\right)=0.
\]
When $r\rightarrow r_{H}$, according to Eq.(\ref{eq:horizonEvap})
, it becomes

\[
-E+q\frac{Q}{r}=2e^{\psi}\dot{r}(m^{2}+\frac{\mathcal{K}}{r^{2}}).
\]

If the evaporation of black hole is very slow, $f\rightarrow0$ (i.e.
$\dot{r}\rightarrow0$), it becomes

\[
-E+q\frac{Q}{r}=0.
\]

(2) $F'=0$, it is

\[
-f'\left(\frac{\mathcal{K}}{r^{2}}+m^{2}\right)+\frac{2\mathcal{K}f}{r^{3}}-2e^{-2\psi}\psi'\left(\frac{qQ}{r}-E\right)^{2}-\frac{2qQe^{-2\psi}\left(\frac{qQ}{r}-E\right)}{r^{2}}=0.
\]
When $F=0$ and $r\rightarrow r_{H}$, it becomes

\[
-f'\left(\frac{\mathcal{K}}{r^{2}}+m^{2}\right)+\frac{2\mathcal{K}2e^{-\psi}\dot{r}}{r^{3}}-8\psi'\left[\dot{r}\left(m^{2}+\frac{\mathcal{K}}{r^{2}}\right)\right]^{2}-\frac{4qQe^{-\psi}\dot{r}\left(m^{2}+\frac{\mathcal{K}}{r^{2}}\right)}{r^{2}}=0.
\]

If the evaporation of black hole is very slow, $f\rightarrow0$ (i.e.
$\dot{r}\rightarrow0$), it is

\[
-f'\left(\frac{\mathcal{K}}{r^{2}}+m^{2}\right)=0.
\]
From Eq.(\ref{hjmetric22}) we know $\mathcal{K}\geq0$. If $m\neq0$
and $m^{2}+\mathcal{K}/r^{2}>0$, we have the description

\[
f'=-\frac{2M'}{r}+\frac{2M}{r^{2}}-\frac{2Q^{2}}{r^{3}}+\frac{2QQ'}{r^{2}}=0.
\]
From this equation we can see that the background black hole does
not have to be an extremal black hole, and it will be extremal only
if $m',Q'\rightarrow0$. In this case it will be nearly a huge stationary
black hole.

(3) $F''=0$, it is

\begin{eqnarray*}
F'' & = & -f''\left(\frac{\mathcal{K}}{r^{2}}+m^{2}\right)+\frac{4\mathcal{K}f'}{r^{3}}-\frac{6\mathcal{K}f}{r^{4}}\\
 &  & +\frac{8qQe^{-2\psi}\psi'\left(\frac{qQ}{r}-E\right)}{r^{2}}\\
 &  & +e^{-2\psi}\left[\frac{2q^{2}Q^{2}}{r^{4}}+\frac{4qQ\left(\frac{qQ}{r}-E\right)}{r^{3}}\right]=0.
\end{eqnarray*}
When $F=0$, $F'=0$ and $r\rightarrow r_{H}$, its expression becomes
very complex. If the evaporation of black hole is very slow $f(r)\rightarrow0$
(i.e. $\dot{r}\rightarrow0$), it becomes

\begin{equation}
-f''\left(\frac{\mathcal{K}}{r^{2}}+m^{2}\right)+e^{-2\psi}\frac{2q^{2}Q^{2}}{r^{4}}=0.\label{eq:tabeEvap}
\end{equation}

In order to simplify the expression, we denote

\[
C(r)=-f''\left(\frac{\mathcal{K}}{r^{2}}+m^{2}\right)+e^{-2\psi}\frac{2q^{2}Q^{2}}{r^{4}},
\]
According to the idea of the article \cite{Harada2011}, we summarize
our result in the table:

\begin{tabular}{|c|c|c|c|}
\hline
Class & $F(r)$ at $r=r_{H}$ & Scenario & Parameter region\tabularnewline
\hline
\hline
I & $F=F'=0,F''>0$ & direct collision & $C(r)>0$\tabularnewline
\hline
II & $F=F'=F''=0$ & LSO collision & $C(r)=0$\tabularnewline
\hline
III & $F=F'=0,F''<0$ & multiple scattering & $C(r)<0$\tabularnewline
\hline
IV & $F=0,F'<0$ & multiple scattering & $f'>0$\tabularnewline
\hline
\end{tabular}

\section{conclusion}

In this paper, we discuss the BSW of slowly evaporating charged black
hole. When the evaporation of blackhole is very slow, the CM energy
is the same as the Reissner-Nordstr\"om black hole near the horizon.
But now the black hole may be not extremal, and it may also be a near-extremal
black hole unless $M'$ and $Q'$ of the black hole are both very
small.

\end{document}